\documentclass[aps,prb,twocolumn,superscriptaddress,nofootinbib,raggedbottom]{revtex4-2}

\usepackage{amsmath}
\usepackage{amssymb}
\usepackage{graphicx}
\makeatletter
\renewcommand{\@dblfloatplacement}{%
  \global\@dbltopnum 1%
  \global\@dbltoproom \dbltopfraction\@colht
  \@textmin \@colht
  \advance \@textmin -\@dbltoproom
  \@fpmin \dblfloatpagefraction\textheight
  \@fptop \@dblfptop
  \@fpsep \@dblfpsep
  \@fpbot \@dblfpbot
  \def\f@depth{1sp}}
\let\orig@flsetnum\@flsetnum
\def\@flsetnum#1{}%
\makeatother

\begin{document}

\title{Strain-tunable van Hove singularity and superconducting dome in monolayer kagome metal CsTi$_3$Bi$_5$}

\author{Run Lv}
\affiliation{Key Laboratory of Materials Physics, Institute of Solid State Physics, HFIPS, Chinese Academy of Sciences, Hefei 230031, China}
\affiliation{University of Science and Technology of China, Hefei 230026, China}

\author{Wenqian Tu}
\affiliation{Key Laboratory of Materials Physics, Institute of Solid State Physics, HFIPS, Chinese Academy of Sciences, Hefei 230031, China}
\affiliation{University of Science and Technology of China, Hefei 230026, China}

\author{Dingfu Shao}
\affiliation{Key Laboratory of Materials Physics, Institute of Solid State Physics, HFIPS, Chinese Academy of Sciences, Hefei 230031, China}

\author{Yuping Sun}
\affiliation{High Magnetic Field Laboratory, HFIPS, Chinese Academy of Sciences, Hefei 230031, China}
\affiliation{Key Laboratory of Materials Physics, Institute of Solid State Physics, HFIPS, Chinese Academy of Sciences, Hefei 230031, China}
\affiliation{Collaborative Innovation Center of Microstructures, Nanjing University, Nanjing 210093, China}

\author{Wenjian Lu}
\email{wjlu@issp.ac.cn}
\affiliation{Key Laboratory of Materials Physics, Institute of Solid State Physics, HFIPS, Chinese Academy of Sciences, Hefei 230031, China}

\begin{abstract}
The recently discovered titanium-based kagome metal CsTi$_3$Bi$_5$, which exhibits superconductivity without an accompanying charge density wave, provides an ideal platform to investigate the intrinsic electronic structure and superconductivity. Moreover, it enables the exploration of superconductivity in the two-dimensional limit of the kagome lattice. Here, we systematically investigate the influence of biaxial mechanical strain ($\varepsilon$) on the electronic structure and superconductivity of monolayer CsTi$_3$Bi$_5$ using first-principles calculations. We demonstrate that the prominent van Hove singularity (VHS) at the M point can be monotonically tuned to the Fermi level by tensile strain. Furthermore, the superconducting transition temperature $T_C$ exhibits a dome-shaped dependence on strain, reaching a maximum of 3.6 K at an optimal tensile strain of $\varepsilon=+2\%$. The average superconducting gap magnitude is 0.35 meV, with an orbital-dependent and anisotropic distribution. Our results provide a detailed understanding of the superconductivity of monolayer CsTi$_3$Bi$_5$ and show that strain is an effective, continuous tuning knob for manipulating the electronic structure and superconductivity.
\end{abstract}
\maketitle

\section{Introduction}

The kagome lattice is well known for its geometric frustration and unique electronic structure. The recently discovered kagome superconductor AV$_3$Sb$_5$ (A=K, Rb, Cs) exhibits exotic quantum orders, including charge density wave (CDW) \cite{zheng2022emergent,liang2021three,ortiz2021fermi}, superconductivity \cite{jiang2023kagome,chen2021double,xu2021multiband}, nematicity \cite{sur2023optimized,nie2022cdw}, pair density wave \cite{chen2021roton,jin2022interplay}, and non-trivial topology \cite{ortiz2020cvsb}. Dimensionality acts as an independent tuning parameter in kagome metals, significantly affecting their charge order and superconductivity. In exfoliated CsV$_3$Sb$_5$, for thicknesses below about 10 layers, the CDW transition temperature increases while the superconducting transition temperature $T_C$ decreases, and the material becomes insulating below 5 layers \cite{song2023anomalous}. The newly discovered titanium-based kagome metal CsTi$_3$Bi$_5$, which exhibits superconductivity without CDW order, offers a clean platform to study intrinsic kagome superconductivity in the two-dimensional limit \cite{yang2024superconductivity,yi2023superconducting}.

CsTi$_3$Bi$_5$ exhibits a distinct electronic structure compared with its vanadium-based counterparts. A notable feature is the van Hove singularity (VHS) just above the Fermi level. In contrast to CsV$_3$Sb$_5$, the VHS in CsTi$_3$Bi$_5$ can be tuned to the Fermi level by chemical doping \cite{liu2021tunable}. However, chemical doping inevitably introduces disorder, so a clean and continuous tuning knob is needed to probe the intrinsic kagome superconductivity. Lattice strain, typically originating from substrate mismatch or mechanical bending, offers such an approach. Strain engineering has indeed proven effective at modulating the electronic structure and superconductivity of diverse two-dimensional materials \cite{qi2023recent,liu2025strain,poon2025superconducting,lin2024uniaxial}. Recently, strain experiments on CsV$_3$Sb$_5$ have shown that its nodal pairing can be enhanced by applying strain \cite{takeuchi2026strained}. Whether strain can effectively tune the electronic structure and superconductivity in the CDW-free kagome metal CsTi$_3$Bi$_5$ remains unexplored.

In this work, we systematically investigate the influence of strain on the electronic structure and superconductivity of monolayer CsTi$_3$Bi$_5$ using first-principles calculations. Our results show that strain can be used as a clean and continuous knob to modulate the electronic structure, Fermi surface topology, and superconductivity of monolayer CsTi$_3$Bi$_5$. We find that as the in-plane lattice parameter expands, the apical Bi atoms approach the kagome plane, similar to the effect of applying pressure in the bulk. The VHS located at the M point can be monotonically tuned to the Fermi level by tensile strain, while compressive strain shifts it to higher energy. A dome-shaped dependence of $T_C$ on strain is observed, with $T_C$ reaching a maximum of 3.6 K at an optimal tensile strain of +2\%. The calculated superconducting gap exhibits a strongly anisotropic and orbital-dependent distribution over the Fermi surfaces, with an average magnitude of 0.35 meV at $\varepsilon=+2\%$. The planar Bi-$6p_z$ orbital yields smaller gap values ($\sim$0.32 meV), while the Ti-$3d_{xz,yz}$ and $3d_{xy,x^{2}-y^{2}}$ orbitals give larger gap values (0.35-0.42 meV). Our results not only provide a detailed understanding of the superconductivity of monolayer CsTi$_3$Bi$_5$ but also show that strain is an effective, continuous tuning knob for manipulating electronic structure and superconductivity.

\section{Methods}

The first-principles calculations based on density functional theory were performed using the Quantum ESPRESSO package \cite{giannozzi2017advanced,giannozzi2009quantum}. Ultrasoft pseudopotentials were used to describe the interaction between electrons and ionic cores \cite{vanderbilt1990soft}. The exchange-correlation interaction was described within the generalized gradient approximation using the Perdew-Burke-Ernzerhof functional \cite{perdew1996generalized}. The cutoffs of kinetic energy and charge density were set to 50 and 500 Ry, respectively. The van der Waals (vdW) interaction was included via Grimme's DFT-D3 method \cite{grimme2010effect}. A vacuum layer of 20 \AA~thickness was introduced to construct the monolayer slab model. The structure was relaxed until the Hellmann-Feynman force acting on each atom was less than $10^{-4}$~Ry/\AA, and the convergence criterion for self-consistent calculations was set to $10^{-8}$~Ry. The Brillouin zone was sampled with 20$\times$20$\times$4 and 5$\times$5$\times$1 k- and q-point grids for electronic and phonon calculations, respectively. The superconducting gaps were calculated using the EPW code \cite{lee2023electron}. The electron-phonon matrix was interpolated onto 40$\times$40$\times$4 k- and q-point grids. The Matsubara frequency cutoff was set to 1 eV, and the Dirac delta functions were replaced by Lorentzians with widths of 100 and 0.5 meV for electrons and phonons, respectively. The Coulomb pseudopotential was set to $\mu^*=0.1$ in both the Allen-Dynes and anisotropic Migdal-Eliashberg calculations.

\section{Results and discussion}

\begin{figure}[!t]
\includegraphics[width=\columnwidth]{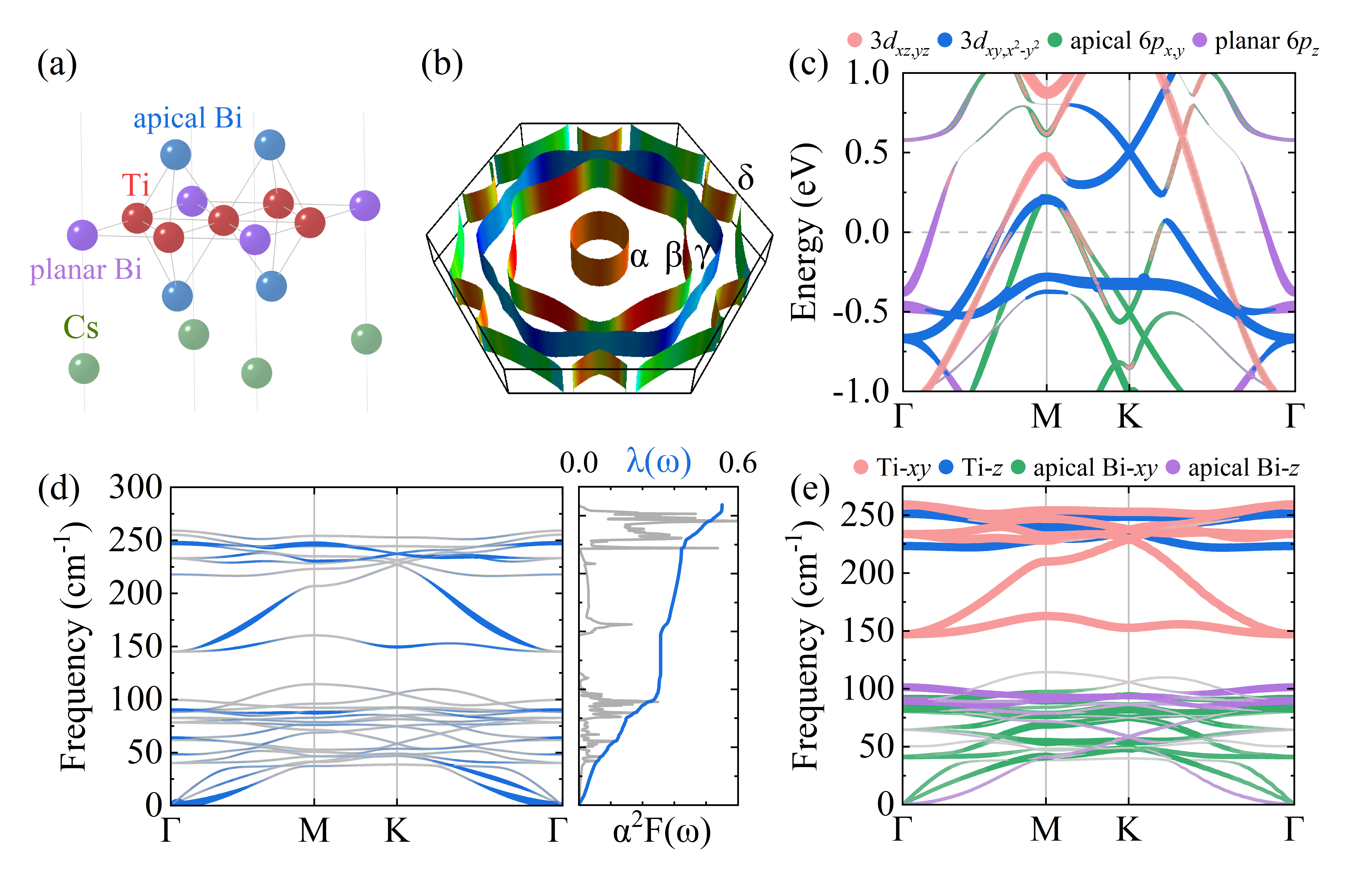}\par
\vspace{6pt}
\caption{(a) Structure of monolayer CsTi$_3$Bi$_5$. (b) Fermi surface. (c) Orbital-resolved band structure. (d) Phonon spectra and integrated electron-phonon coupling (EPC) strength $\lambda$. The width of the blue lines indicates the mode-resolved EPC strength $\lambda_{\mathbf{q}\nu}$. (e) Phonon spectra projected onto the vibrational modes of Ti and apical Bi atoms.}
\label{fig:fig1}
\end{figure}

The titanium-based kagome metal CsTi$_3$Bi$_5$ exhibits a layered structure in the $P6/mmm$ space group, identical to that of the vanadium-based kagome superconductors of the AV$_3$Sb$_5$ family \cite{yang2024superconductivity}. Figure~\ref{fig:fig1}(a) shows the structure of monolayer CsTi$_3$Bi$_5$, where the kagome lattice is formed by Ti and planar Bi atoms. The Bi atoms in the kagome plane and those outside it are denoted as planar and apical Bi, respectively. The band structure is also similar to that of CsV$_3$Sb$_5$, but its Fermi energy ($E_F$) is lower (see Fig.~S1). We begin by investigating the unstrained structure to provide a baseline for understanding strain effects on the electronic structure and superconductivity of monolayer CsTi$_3$Bi$_5$. Figure~\ref{fig:fig1}(c) shows the orbital-resolved band structure. Four bands cross the Fermi level, originating from the Ti-$3d_{xz,yz}$, Ti-$3d_{xy,x^{2}-y^{2}}$, apical Bi-$6p_{x,y}$ and planar Bi-$6p_z$ orbitals. The $\alpha$ band close to the $\Gamma$ point originates from the Bi-$6p_z$ orbital, forming a cylindrical Fermi surface (FS). The hexagonal $\beta$ and $\gamma$ FSs are primarily composed of the Ti-$3d_{xz,yz}$ and $3d_{xy,x^{2}-y^{2}}$ orbitals (see Fig.~S2), respectively, with a minor $6p_{x,y}$ contribution from apical Bi to both sheets (see Fig.~S3(d)). The $\delta$ FS close to the M and K points is mainly derived from Ti-$3d_{xz,yz}$ orbitals.

In the unstrained structure, the calculated total EPC strength $\lambda$ is 0.54, as shown in the right panel of Fig.~\ref{fig:fig1}(d). The width of the blue lines on the phonon spectra indicates the mode-resolved EPC strength ($\lambda_{\mathbf{q}\nu}$), calculated by
\begin{multline}
\lambda_{\mathbf{q}\nu}=\frac{2}{\hbar N E_{F}\omega_{\mathbf{q}\nu}}%
\times\sum_{ij}\int_{BZ}\frac{dk}{\Omega_{BZ}}\\
\left|g_{\mathbf{q}\nu}(k,i,j)\right|^{2}%
\delta(\epsilon_{k,i}-E_{F})\,\delta(\epsilon_{k+\mathbf{q},j}-E_{F}),
\label{eq:eplambda}
\end{multline}
where $g_{\mathbf{q}\nu}$ stands for the EPC matrix elements. The phonon modes with large $\lambda_{\mathbf{q}\nu}$ originate primarily from the low-frequency out-of-plane vibrations of the apical Bi atoms and the high-frequency in-plane vibrations of the Ti atoms (see Fig.~S4 for all atomic contributions). Based on the Allen-Dynes modified McMillan equation
\begin{equation}
T_{c}=\frac{\omega_{\mathrm{log}}}{1.2}\exp\left[-\frac{1.04(1+\lambda)}{\lambda-\mu^{*}-0.62\,\lambda\mu^{*}}\right],
\label{eq:mcmillan}
\end{equation}
the estimated superconducting transition temperature ($T_C$) is 2.6 K.

\begin{figure}[!t]
\includegraphics[width=\columnwidth]{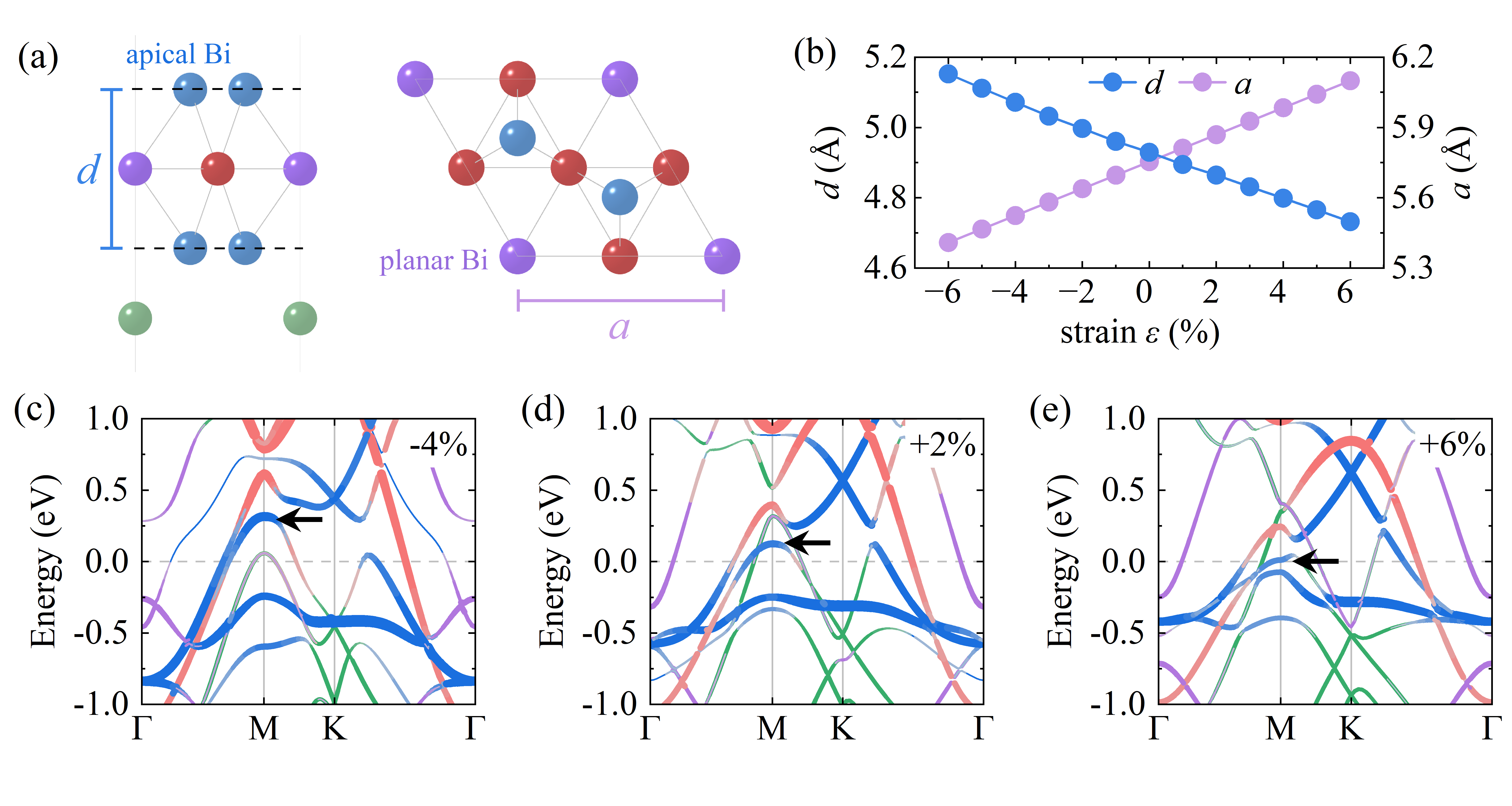}\par
\vspace{6pt}
\caption{(a) Lateral and top views of monolayer CsTi$_3$Bi$_5$. (b) Changes in the distance between apical Bi atoms. (c)-(e) Evolution of the band structure with applied strain $\varepsilon$. The black arrow highlights the VHS located at the M point.}
\label{fig:fig2}
\end{figure}

To investigate the influence of strain on the electronic structure and superconductivity, we apply biaxial strain in the ab plane by simultaneously adjusting the lattice parameters $a$ and $b$. The strain magnitude is defined as $\varepsilon=(a-a_0)/a_0\times100\%$, where $a_0$ is the unstrained lattice constant. Positive (negative) values of $\varepsilon$ thus denote tensile (compressive) strain. The in-plane lattice parameters are fixed according to the applied strain $\varepsilon$, while the out-of-plane z-coordinates of the apical Bi and Cs atoms are fully relaxed to minimize the total energy. Since the Cs atoms move along z in the same way as the neighboring apical Bi atoms and contribute negligibly to the total EPC strength, we plot only the distance $d$ between the two apical Bi atoms in Fig.~\ref{fig:fig2}(b). Increasing the strain $\varepsilon$ monotonically drives the apical Bi atoms toward the kagome plane, an effect analogous to that of hydrostatic pressure on the bulk crystal. In addition, this biaxial distortion enlarges the lattice parameter $a$, thus expanding the kagome plane and increasing the distance between Ti atoms. Mechanical strain therefore produces effects that are partly analogous to, and partly distinct from, those of hydrostatic pressure.

We calculate the band structures under various strain conditions, ranging from compressive ($\varepsilon=-4\%$) to tensile ($\varepsilon=+6\%$). The unstrained band structure ($\varepsilon=0\%$) shows the VHS above $E_F$ at the M point, composed of the Ti-$3d_{xy,x^{2}-y^{2}}$ orbitals (blue lines). Under compressive strain (Fig.~\ref{fig:fig2}(c)), the VHS shifts to higher energy, moving further from the Fermi level. The hole-like band around the M point, which arises from apical Bi-$6p_{x,y}$ and Ti-$3d_{xz,yz}$ orbitals, shifts to lower energy. Conversely, tensile strain (Figs.~\ref{fig:fig2}(d) and (e)) causes the VHS to shift toward lower energy. At $\varepsilon=+6\%$, the VHS is tuned precisely to the Fermi level. The flat band below $E_F$ likewise approaches the Fermi level. This monotonic, strain-driven shift of the VHS at the M point to the Fermi level (highlighted by black arrows) demonstrates that the electronic structure of this kagome metal can be manipulated by mechanical strain. Such a strain-tunable VHS offers a route to probing its relationship with competing electronic orders.

\begin{figure}[!t]
\includegraphics[width=\columnwidth]{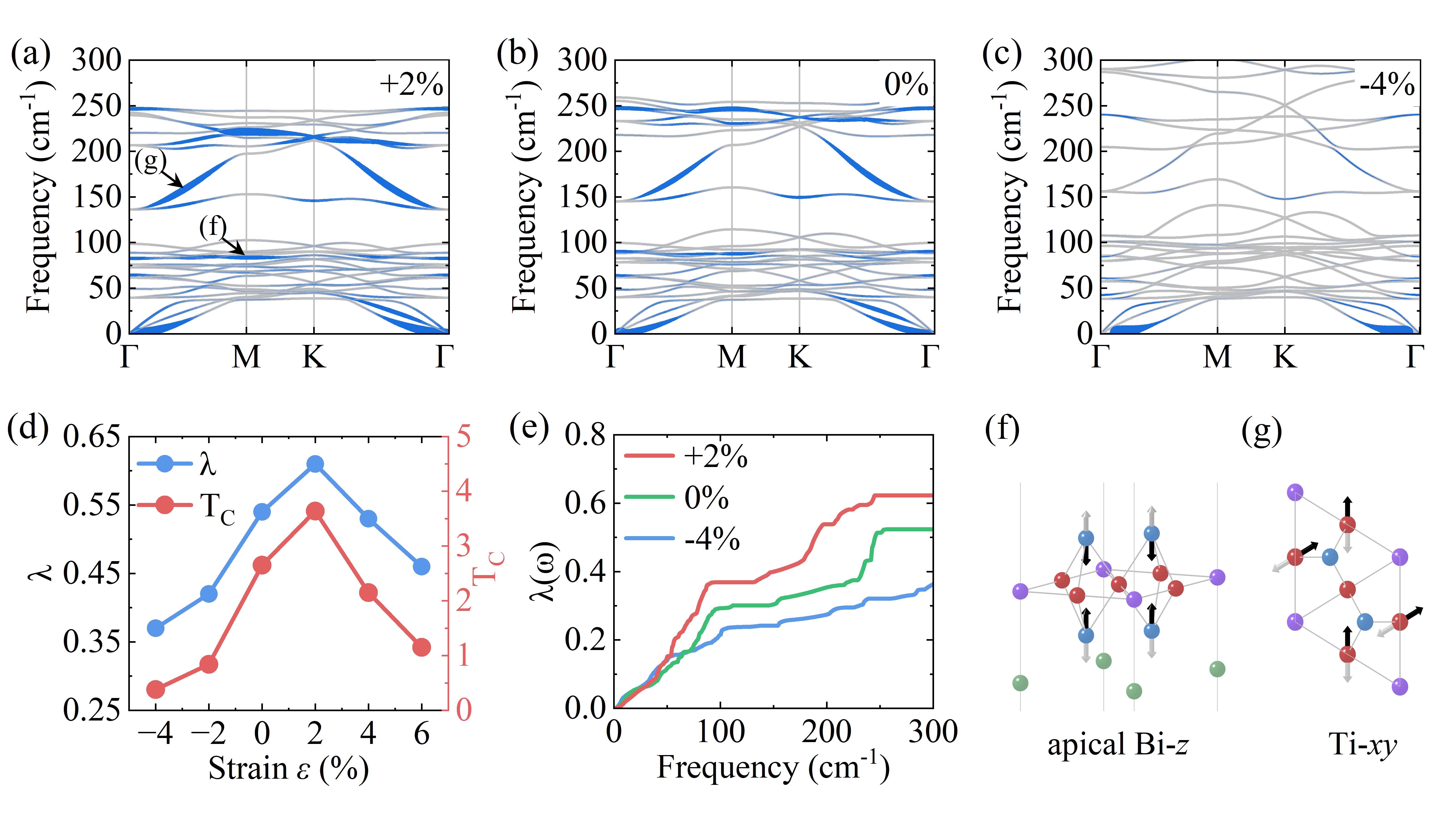}\par
\vspace{6pt}
\caption{(a)-(c) Evolution of the phonon spectra with applied strain $\varepsilon$. The width of blue lines indicates EPC strength $\lambda_{\mathbf{q}\nu}$. (d) Total EPC strength $\lambda$ and superconducting transition temperature $T_C$ versus strain $\varepsilon$. (e) Frequency-integrated EPC strength $\lambda(\omega)$. (f)-(g) Vibrational schematics of corresponding phonon modes.}
\label{fig:fig3}
\end{figure}

To elucidate the influence of strain on the superconductivity of monolayer CsTi$_3$Bi$_5$, we calculate the phonon spectra and the associated electron-phonon coupling for various strain values. Figures~\ref{fig:fig3}(a)-(c) present the calculated phonon dispersions, showing that the phonon modes soften as the strain increases from compressive to tensile values. The mode- and momentum-resolved EPC strength ($\lambda_{\mathbf{q}\nu}$) is represented by the width of the blue lines. As strain increases from -4\% to +2\%, the mode-resolved EPC strength $\lambda_{\mathbf{q}\nu}$ of the two modes marked by black arrows increases (Fig.~\ref{fig:fig3}(a)). These two modes are dominated by the out-of-plane vibrations of the apical Bi atoms (Fig.~\ref{fig:fig3}(f)) and the in-plane vibrations of the Ti atoms (Fig.~\ref{fig:fig3}(g)), respectively. Integration of the Eliashberg spectral function $\alpha^{2}F(\omega)$ gives the total EPC strength $\lambda$, which increases from 0.37 at $\varepsilon=-4\%$ to 0.62 at $\varepsilon=+2\%$ (Fig.~\ref{fig:fig3}(e)). Beyond +2\%, $\lambda$ decreases to 0.46 at $\varepsilon=+6\%$. The nonmonotonic strain dependence of EPC, which first increases and then decreases, gives rise to a dome-shaped dependence of $T_C$ on strain. As summarized in Fig.~\ref{fig:fig3}(d), $T_C$, estimated using the Allen-Dynes modified McMillan equation, rises from 0.3 K at -4\% to a maximum of 3.6 K at +2\%, and then falls to 1.2 K at +6\%.

\begin{figure}[!t]
\includegraphics[width=\columnwidth]{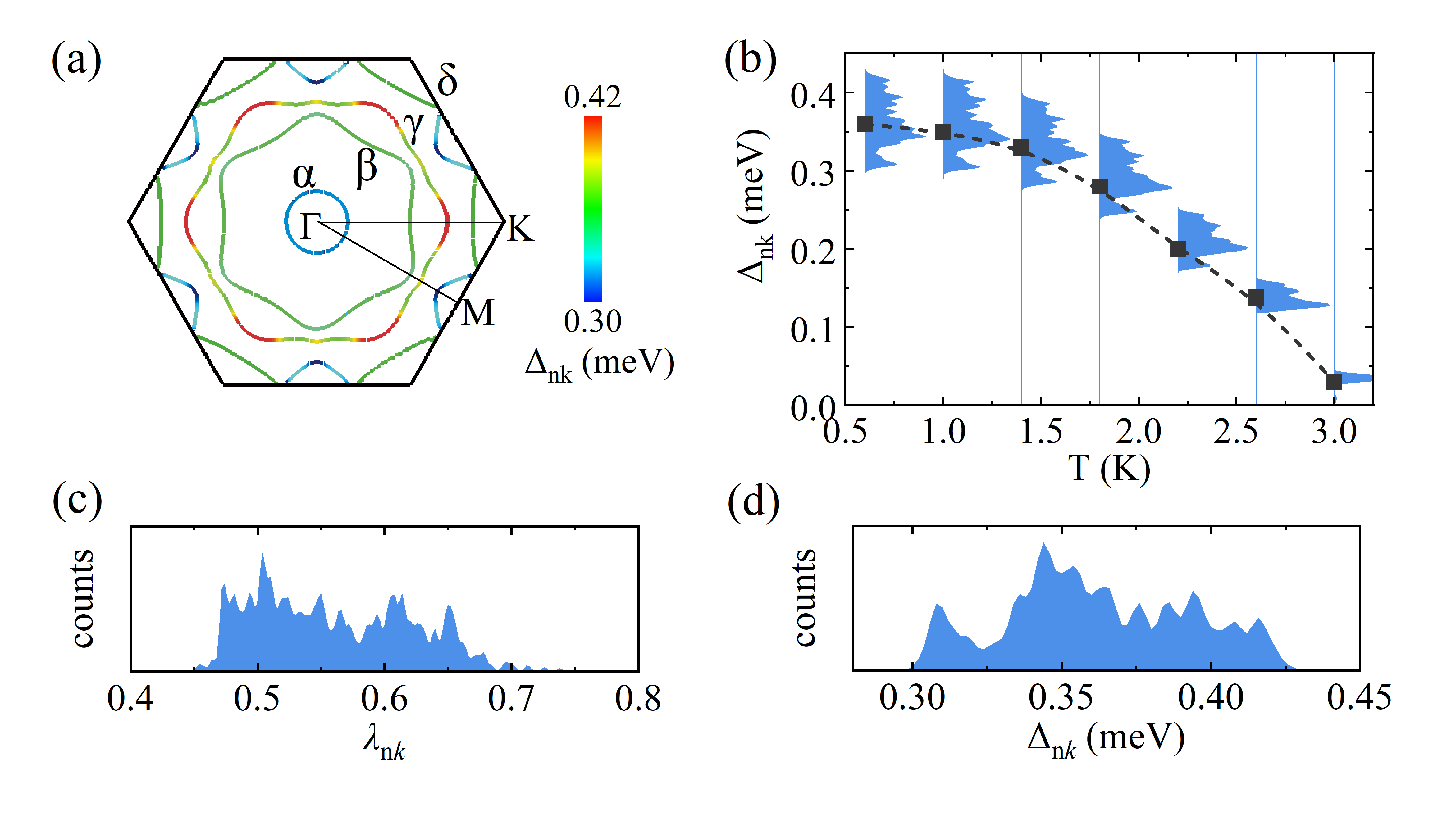}\par
\vspace{6pt}
\caption{(a) Momentum-resolved superconducting gap $\Delta_{n\mathbf{k}}$ on the Fermi surface, calculated by anisotropic Migdal-Eliashberg theory. (b) Energy distribution of the superconducting gap $\Delta_{n\mathbf{k}}$ as a function of temperature. (c), (d) Distribution of EPC strength $\lambda_{n\mathbf{k}}$ and superconducting gap $\Delta_{n\mathbf{k}}$ around the Fermi surface.}
\label{fig:fig4}
\end{figure}

To gain a deeper insight into the superconductivity in monolayer CsTi$_3$Bi$_5$, we calculate the momentum-resolved superconducting gap $\Delta_{n\mathbf{k}}$ using the anisotropic Migdal-Eliashberg theory. Figure~\ref{fig:fig4}(a) shows the distribution of the superconducting gap on the Fermi surface under the optimal tensile strain ($\varepsilon=+2\%$). The gap is fully open, with an average magnitude of 0.35 meV. The central pocket is dominated by the planar Bi-$6p_z$ orbital, exhibiting the smallest gap magnitude of 0.32 meV. The hexagonal Fermi surfaces $\beta$ and $\gamma$ are dominated by Ti-$3d_{xz,yz}$ and $3d_{xy,x^{2}-y^{2}}$ orbitals, respectively, with gap values ranging from 0.35 to 0.42 meV. The sheet of $\delta$ FS around the K point exhibits moderate gap magnitudes and also mainly arises from Ti-$3d_{xz,yz}$ orbitals. Solving the anisotropic Migdal-Eliashberg equations at different temperatures gives the temperature dependence of $\Delta_{n\mathbf{k}}$, from which we estimate $T_C=3.2$ K, close to the $T_C$ estimated by the Allen-Dynes modified McMillan equation.

Kagome metals such as the AV$_3$Sb$_5$ family have attracted intense interest in their superconductivity. The recently discovered CsTi$_3$Bi$_5$ has been reported to exhibit superconductivity without structural instability, thereby providing an ideal system to access the intrinsic superconductivity of the kagome lattice. Our first-principles calculations systematically demonstrate that biaxial strain is an effective and clean tool to manipulate the electronic structure and significantly enhance the electron-phonon-mediated superconductivity in monolayer CsTi$_3$Bi$_5$. The VHS at the M point can be continuously tuned to the Fermi energy by tensile strain, culminating in a precise alignment at $\varepsilon=+6\%$. Furthermore, $T_C$ can be enhanced to 3.6 K at the optimal strain $\varepsilon=+2\%$. This value is higher than the ambient-pressure bulk $T_C$ of 1.85 K estimated within the first-principles framework \cite{yi2023superconducting}. Notably, the VHS reaches the Fermi level at $\varepsilon=+6\%$, whereas $T_C$ is maximized at $\varepsilon=+2\%$. This indicates that the $T_C$ enhancement is not determined solely by the VHS-$E_F$ alignment but also by the strain dependence of $\lambda$. Finally, the resulting orbital dependence of the gap resembles the behavior inferred for CsV$_3$Sb$_5$, suggesting a common origin across titanium- and vanadium-based kagome superconductors.

\section{Conclusion}

In conclusion, through first-principles calculations, we have systematically investigated the influence of biaxial strain on the electronic structure and superconductivity of monolayer CsTi$_3$Bi$_5$. Notably, the VHS at the M point can be monotonically tuned to the Fermi level by applying tensile strain. In addition, tensile strain can significantly enhance $T_C$ from 0.3 K to 3.6 K as the strain varies from -4\% to +2\%, and then falls to 1.2 K at +6\%, giving a dome-shaped dependence of $T_C$ on strain. Within the anisotropic Migdal-Eliashberg framework, the calculated superconducting gap attains an average magnitude of 0.35 meV at $\varepsilon=+2\%$ and exhibits an orbital-dependent and anisotropic distribution. More generally, these results position biaxial strain as a disorder-free and continuously tunable knob for exploring how lattice geometry controls pairing in two-dimensional kagome metals.

\section*{Acknowledgments}
This work was supported by the National Key Research and Development Program of China under Contract No.\ 2022YFA1403203.

\bibliographystyle{apsrev4-2}
\bibliography{refs}

\end{document}